\documentclass[twoside,twocolumn,9pt]{article}

\usepackage{graphicx}
\usepackage{dcolumn}
\usepackage{bm}
\usepackage{color}
\usepackage{amsmath}
\usepackage{amssymb}
\usepackage[compact]{titlesec}
\usepackage[font=small]{caption}
\usepackage{array}
\usepackage{multirow}
\usepackage{adjustbox}
\usepackage{setspace}
\usepackage[T1]{fontenc}
\usepackage[outdir=./]{epstopdf}
\usepackage[super,sort&compress,comma]{natbib} 
\usepackage{notoccite}
\usepackage[left=1.5cm, right=1.5cm, top=1.785cm, bottom=2.0cm]{geometry}
\usepackage{authblk}
\usepackage{sectsty}
\usepackage{charter}
\usepackage[version=3]{mhchem}
\usepackage{balance}
\usepackage{abstract}
\usepackage{fancyhdr}
\usepackage{fnpos}
\usepackage{array}
\usepackage{floatrow}
\usepackage{subfig}
\usepackage{comment}
\usepackage{enumitem}
\newlength\mysep 
\DeclareFloatSeparators{mysep}{\hskip\mysep}
\usepackage{dblfloatfix}    

\begin{document}

\makeFNbottom
\renewcommand{\thefootnote}{\fnsymbol{footnote}}
\setlength{\parskip}{0pt}
\setlength\topsep{0pt plus 0pt minus 0pt}
\setlength{\textfloatsep}{2pt plus 2pt minus 2pt}
\setlength{\floatsep}{2pt plus 2pt minus 2pt}
\graphicspath {{Figures/}}
\titlespacing*{\section}
{0pt}{5ex plus 1ex minus .2ex}{4.3ex plus .2ex}
\titlespacing*{\subsection}
{0pt}{5ex plus 1ex minus .2ex}{4.3ex plus .2ex}
\newcolumntype{P}[1]{>{\centering\arraybackslash}p{#1}}

\renewcommand{\abstractname}{}    
\renewcommand{\absnamepos}{empty} 

\twocolumn[
\begin{@twocolumnfalse}
\centering
\LARGE{\textbf{Automated Electrokinetic Stretcher for Manipulating Nanomaterials}} \\ \vspace{0.5cm}
\large{Beatrice W. Soh,\textit{$^{a*}$} Zi-En Ooi,\textit{$^{a*}$} Eleonore Vissol-Gaudin,\textit{$^b$} Chang Jie Leong,\textit{$^a$} and Kedar Hippalgaonkar\textit{$^{a,b}$}} \\ \vspace{0.8cm}

\begin{abstract}
In this work, we present an automated platform for trapping and stretching individual micro- and nanoscale objects in solution using electrokinetic forces. The platform can trap objects at the stagnation point of a planar elongational electrokinetic field for long time scales, as demonstrated by the trapping of $\sim 100$ nm polystyrene beads and DNA molecules for minutes, with a standard deviation
in displacement from the trap center $<1 \ \mu$m. This capability enables the stretching of deformable nanoscale objects in a high-throughput fashion, as illustrated by the stretching of more than 400 DNA molecules within $\sim$4 hours. The flexibility of the electrokinetic stretcher opens up numerous possibilities for contact-free manipulation, with size-based sorting of DNA molecules performed as an example. The platform described provides an automated, high-throughput method to track and manipulate objects for real-time studies of micro- and nanoscale systems.
\end{abstract}
\end{@twocolumnfalse}
]

\footnotetext{\textit{$^{a}$~Institute of Materials Research and Engineering (IMRE), Agency for Science, Technology and Research, Singapore (A*STAR). E-mail: beatrice\_soh@imre.a-star.edu.sg and kedar@ntu.edu.sg}}
\footnotetext{\textit{$^{b}$~Department of Materials Science and Engineering, Nanyang Technological University, Singapore}}
\footnotetext{$^*$B.W. Soh and Z.-E. Ooi contributed equally to this work.}

\section{Introduction}
The trapping and manipulation of small objects is an important tool in a broad range of scientific fields, from nanofabrication\cite{grier2003revolution,urban2014optical} to single-molecule biophysics.\cite{moerner2007new,bustamante2021optical,gross2011quantifying} An array of techniques have been developed over the past few decades to trap and manipulate single objects on the micro- and nanoscale in solution, based on various methods including hydrodynamic,\cite{perkins1997single,smith1998response,tanyeri2010hydrodynamic} electrokinetic,\cite{cohen2005method,armani2006using,klotz2018motion,soh2020deformation} optical\cite{ashkin1986observation,neuman2004optical} and magnetic fields.\cite{yan2004near,lee2004manipulation} The ability to trap and manipulate micro- and nano-objects has led to major strides particularly in the real-time study of biological systems.\cite{probst2012flow,bradac2018nanoscale,zhang2021plasmonic}

The advantage of using the stagnation point of fluid flows to trap and stretch molecules and particles is the noncontact and minimally invasive nature of the approach, which does not require tethering.\cite{brimmo2017stagnation} It has been employed for applications such as droplet characterization \cite{bentley1986computer,narayan2020droplet} and DNA stretching.\cite{perkins1997single,smith1998response,dylla2010single,soh2020deformation} While single DNA molecules have been trapped for long-time observation using planar elongational electric fields, such studies were performed via manual control of the trapping fields.\cite{klepinger2015stretching,soh2020deformation,perkins1997single,smith1998response} 
Single molecule stretching and motion control is also an important capability particularly for DNA sequencing techniques, with methods in the literature based primarily on flow through nanoscale channels.\cite{zrehen2019chip,niedzwiecki2021devices,lam2012genome}
In implementing an automated platform that simultaneously traps \textit{and} stretches a molecule, the development of a robust tracking system is a major challenge. Given the continually evolving molecule shape in time, a means of tracking the object position (e.g. center-of-mass) is needed.

Herein, we report an automated, high-throughput electrokinetic stretcher for the observation and manipulation of single objects on the micro- and nanoscale in solution. The platform can operate as a trap that implements active feedback control to maintain the center-of-mass of a deformable object at the stagnation point of a planar elongational electrokinetic field, generated in a microfluidic cross-slot device. We first demonstrate the trapping of submicron particles and double-stranded DNA molecules for extended periods of time without viscosity modification. Then, we show the platform's ability to trap and stretch DNA molecules in a self-driven, high-throughput manner, which can be used to obtain large datasets of stretching trajectories. Lastly, we also showcase the flexibility of the platform to perform other types of manipulation, demonstrating specifically its application for sorting molecules by size. 

\section{Working principle}

The electrokinetic trap consists of a 2 $\mu$m tall polydimethylsiloxane (PDMS) channel with a cross-slot geometry sealed to a glass slide, as presented in Fig. \ref{fig:schematic}. Potentials independently applied to the four reservoirs of the cross-slot channel generate a planar elongational electric field, the kinematics of which is given by:
\begin{align}
    \dot{x} &= \mu E_x = \dot{\epsilon}x  \\
    \dot{y} &= \mu E_y = -\dot{\epsilon}y
\end{align}
where $\dot{x}$ and $\dot{y}$ are the velocities in the $x$ and $y$ directions respectively, $E_x$ and $E_y$ are the electric fields in the $x$ and $y$ directions respectively, $\mu$ is the electrokinetic mobility, and $\dot{\epsilon}$ is the strain rate. Electrokinetic motion often arises from both electrophoresis and electroosmosis, and their net effect can be considered together phenomenologically under a single mobility.\cite{Probst2012FlowControl} As such, both charged and uncharged objects can be trapped.\cite{probst2012}

A planar elongational field is described by an axis of pure elongation ($x$) and an orthogonal axis of pure compression ($y$), with a stagnation point at the center of the channel where the local velocity is zero. The stagnation point is a saddle point, with a potential well in the direction of the compressional axis and a potential hill along the elongational axis. If unperturbed, an object can remain at the stagnation point for an indefinite amount of time. However, without a control field, random Brownian collisions inevitably push it away from the stagnation point and cause it to accelerate away along the elongational axis. The potential well in the compressional axis provides passive trapping; hence trapping an object in both dimensions requires only active feedback control along the elongational axis. Our feedback control scheme aims to maintain the center-of-mass of the object at the stagnation point (Fig. \ref{fig:schematic}(a))

\section{Platform implementation}

\subsection{Channel and materials preparation}

\begin{figure}[tb!]
\includegraphics[width = 0.92\textwidth]{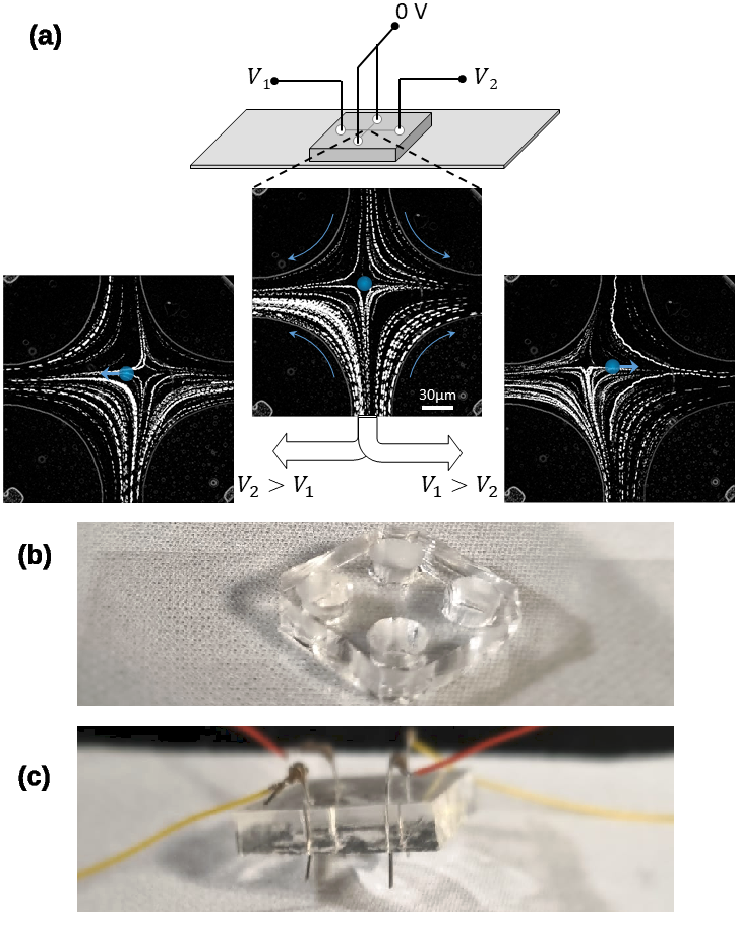}
\caption{\label{fig:schematic}Schematic of the electrokinetic stretcher and its physical implementation in PDMS. (a) Top: Schematic of the microfluidic device and the arrangement of electrodes in the reservoirs where $V_1$, $V_2$ are computer-controlled voltages. Center: Extended-exposure fluorescence image showing traces of DNA molecules flowing through the cross-slot channel symmetrically. The blue circle represents an object at the stagnation point. Left, right: $V_1 \neq V_2$ results in an asymmetric field, which can be used to trap and/or manipulate an object in the presence of an elongational field. (b) Image of PDMS microfluidic cross-slot channel. (c) Reusable PDMS/platinum cover-electrode set-up: improves the mechanical stability of the set-up and reduces evaporation, thus extending sample lifespan up to a week.}
\end{figure}

To prepare the microfluidic device shown in Fig. \ref{fig:schematic}(b), PDMS (Sylgard 184, Dow Corning) was mixed in a 10:1 elastomer/curing agent ratio and cured over a patterned \makebox{4-inch} silicon wafer for 3 hours at 70$^{\circ}$C. The cured PDMS was cut into individual channels and the reservoirs were punched with a 5 mm biopsy punch, followed by rinsing and sonication in ethanol. The glass slides were cleaned with ethanol and soaked in 1M NaOH solution for 1 hour. Finally, the PDMS channels and glass slides were rinsed with water, dried with nitrogen gas and assembled together.

A PDMS-based electrode-cover set-up (Fig. \ref{fig:schematic}(c)) was prepared and placed atop the microfluidic channel. The set-up has two functions: first, the cover acts to stabilize the electrodes inserted into the reservoirs of the microfluidic channel; second, covering the reservoirs helps to minimize buffer evaporation during experiments, so as to maximize device lifetime. A flat PDMS slab was cut into squares that ensure full coverage of the four wells of the microfluidic device. The four electrodes ($\sim2$ cm long platinum wires) were pierced vertically through the PDMS cover until $\sim0.5$ cm of wire protruded from the bottom of the cover to enable contact with the buffer. The top of the platinum electrodes were connected to four copper wires using silver paste. 

In a typical experiment, the buffer consists of 0.5X Tris-boric acid-EDTA (TBE, AccuGENE) solution, polyvinylpyrrolidone (Sigma-Aldrich) and 2-mercaptoethanol (Sigma-Aldrich). The fluorescent polystyrene beads (Bangs Laboratories and Merck Chimie S.A.S) were diluted to $\sim 0.001\%$ solids concentration prior to further dilution in the buffer. T4 DNA (165.6 kbp, Nippon Gene) and $\lambda$ DNA (48.5 kbp, Sigma-Aldrich) were stained with fluorescent dye YOYO-1
(Invitrogen) at a base pair to dye ratio of 4:1, before dilution with the buffer to an appropriate viewing concentration.

\subsection{Platform hardware}

The microfluidic device is viewed through an inverted optical fluorescence microscope (Nikon Eclipse Ti2-U) under 470 nm LED excitation and 60X oil-immersion objective. A dichroic filter is used to reject excitation light from the fluorescence image, which is captured on a Photometrics Prime 95B sCMOS camera and subsequently transferred to a computer for processing. To reduce background noise and minimize light dosage, the LED duty cycle is synchronized to the camera exposure cycle. Platinum wires are immersed into each of the four fluid reservoirs as electrodes, of which two are grounded. The other two electrodes each receive an amplified voltage from a computer-controlled data acquisition (DAQ) board. The maximum slew rate of the amplified voltage was measured to be about 60 V/${\mu}$s. 

\begin{figure}[tb!]
\includegraphics[width = 0.92\textwidth]{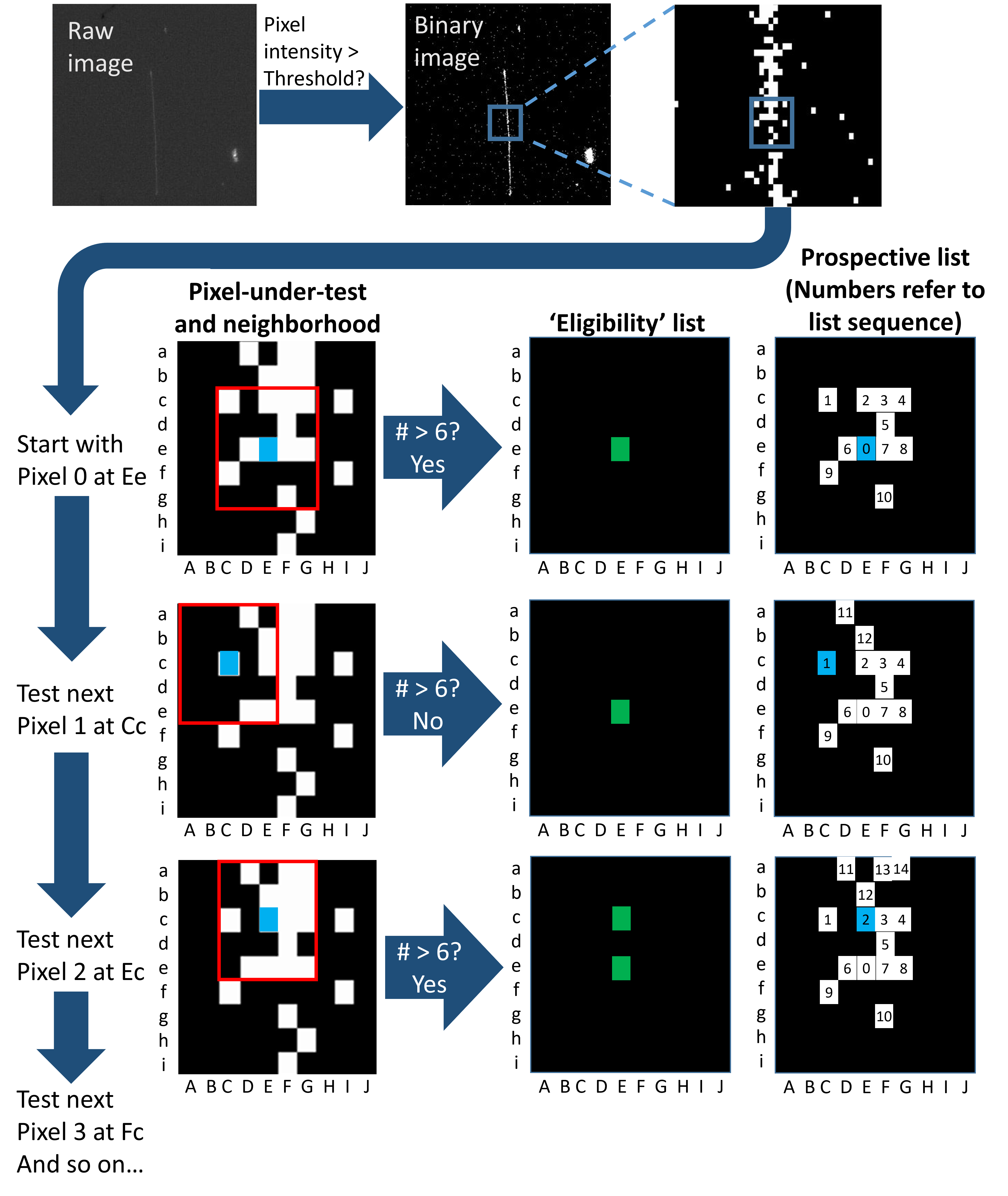}
\caption{\label{fig:densityclustering}Visual illustration of the density-based clustering algorithm developed and implemented to track deformable objects, using a 5-by-5 pixel neighborhood (highlighted by red square) and minimum pixel density of 6. The algorithm is effective at identifying the pixels that belong to the trapped molecule under observation and eliminating the pixels that belong to other molecules or noise.}
\end{figure}

\subsection{Tracking of deformable objects}
A custom LabVIEW program was written for hardware control and image processing. During the initialization phase of the system, the program captures a long-exposure image of free-moving particles at some nonzero voltage to determine the pixel coordinates of the stagnation point. The program then automatically searches the camera frame for an object to trap, prioritizing the object that is closest to the stagnation point. A central horizontal zone is avoided to prevent the re-capture of molecules that have been intentionally released from the trap. Once the system detects an object, a small tracking region-of-interest (ROI) is defined around the particle, and all processing is performed on this ROI to minimize system latency. The program synchronizes image processing to the image capture cycle of the camera, which yields a well-defined latency of exactly one frame time. Minimum latency is limited by image transfer and processing speed to about 10 ms for small, brightly fluorescent objects; extended, dimly fluorescent objects (such as DNA) require longer processing and exposure times, typically 50 ms. 

The electrokinetic stretcher is based on an elongational field, hence elastic objects can deform during the trapping process. To continuously track deformable nanoscale objects, such as a stretched DNA molecule, we implement a density-based clustering algorithm (Fig. \ref{fig:densityclustering}). Based on a selected threshold, the image is first converted into a binary image as input to the algorithm:

\begin{enumerate}
\item The algorithm begins with the first bright pixel closest to the stagnation point.
\item This bright pixel is tested for a minimum number of other bright pixels in its neighborhood. If this minimum density is met, then the pixel-under-test (PUT) is added to a list of eligible pixels. Simultaneously, all other bright pixels in its neighborhood are added to a separate list of prospective pixels to be tested.
\item The next pixel in the prospective list is tested for the minimum density and the process of growing or not growing the two lists repeats.
\item The algorithm terminates when the prospective list stops growing and all prospective pixels have been tested.
\end{enumerate}

The output of the clustering algorithm is a list of pixels that have met the minimum neighborhood density and are contiguous with the starting pixel. This list is applied to the original image to obtain a new image that excludes both noise pixels and pixels that belong to other, interfering molecules. The speed of the clustering algorithm depends more strongly on the area extent of the molecule rather than the size of the image. On a desktop computer running an Intel i7-10700K, we find the clustering can be completed within 4 ms for a cluster containing 500-600 pixels. Although the clustering algorithm was specifically created to track extended, deformable objects, it works equally well for rigid, point-like particles such as polystyrene beads. The algorithm performs even faster because such objects often present as smaller clusters.

The new, filtered image resulting from the clustering algorithm is used to determine the position of the object by calculating its intensity centroid. The offset of the centroid from the stagnation point in the elongational axis is then converted to a linearly proportional voltage correction, tilting the attractive potentials $V_1$ and $V_2$ (Fig. \ref{fig:schematic}). Typical values of this proportional gain range between 5 and 50 V/$\mu$m. 

\section{Results}

\subsection{Trapping capabilities}

\begin{figure*}[tb]
\begin{center}
\includegraphics[width = 0.75\textwidth]{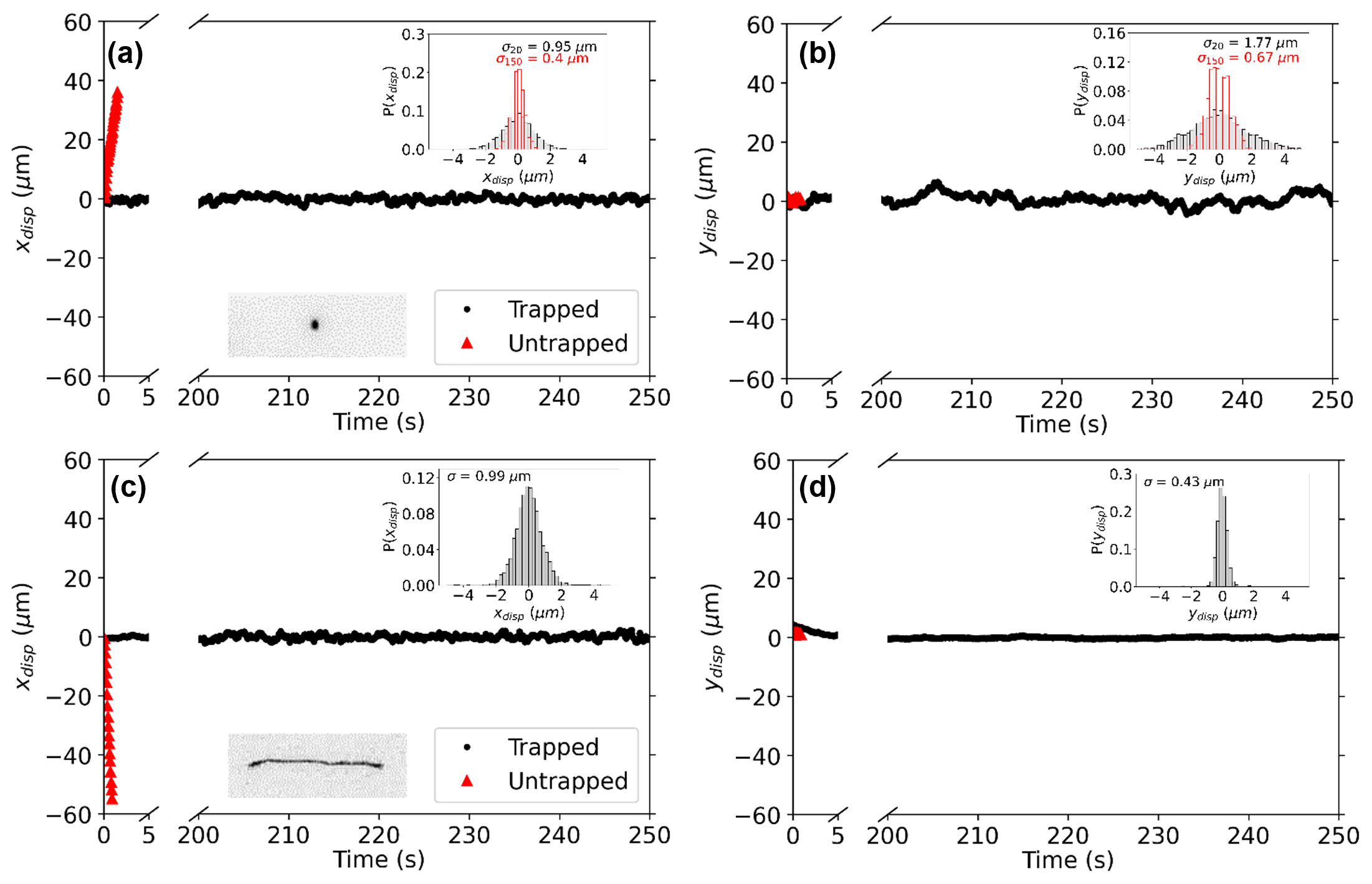}
\end{center}
  \caption{Trapping of nanoscale objects. Trapping of a single (a,b) 190 nm polystyrene bead and (c,d) 165.6 kbp DNA molecule using a 20 V stretching voltage and proportional gain setting of 1.1 and 2.2 V/$\mu$m respectively. Comparison of trajectories of the objects along the elongational ($x$) axis (a,c) and compressional ($y$) axis (b,d) when trapped (black) and not trapped (red). For clarity, the images' intensity gradients have been inverted. Inset: Histogram of displacements of the trapped object from the trap center along the elongational ($x$) and compressional ($y$) axes (grey). Histogram of displacements of a trapped 190 nm polystyrene bead using a 150 V stretching voltage and proportional gain setting of 13.3 V/$\mu$m, in red for comparison, demonstrates much tighter particle confinement.}\label{fig:trap}
\end{figure*}

\begin{figure}[tb]
\includegraphics[width=\textwidth]{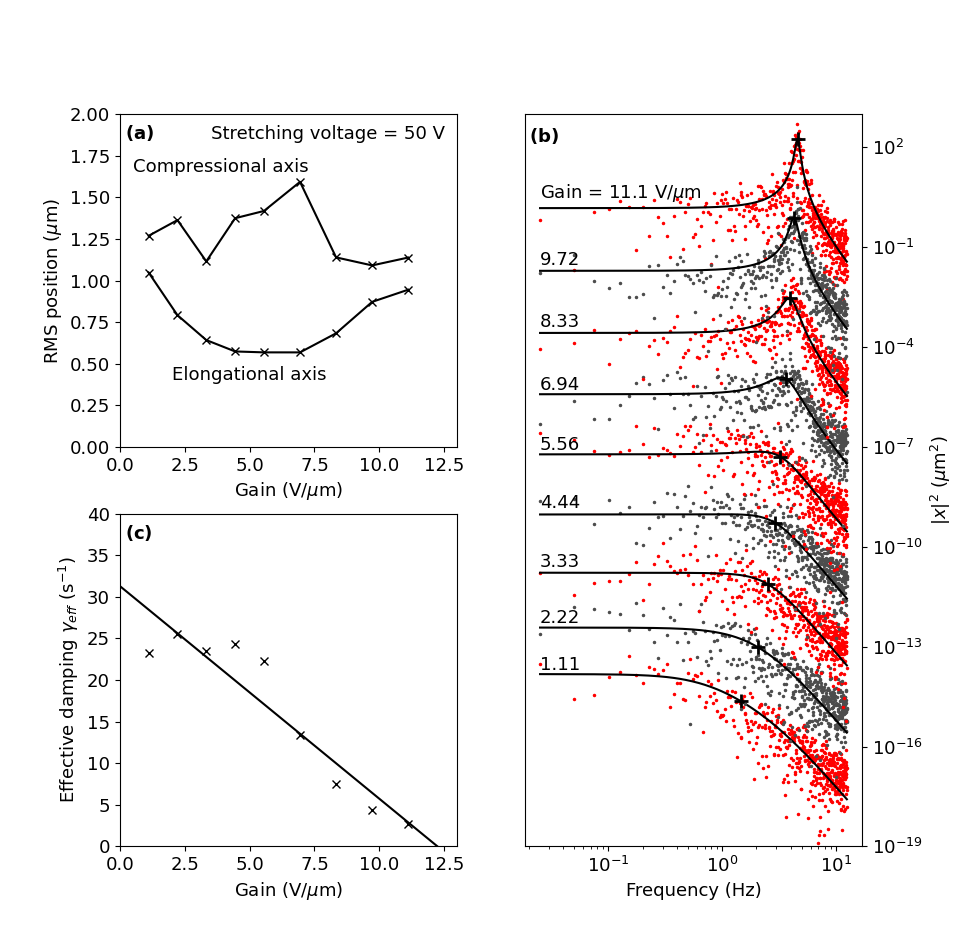}
\caption{\label{fig:freqresp} Varying proportional gain value at trap voltage of 50 V. (a) Positional RMS in both axes as a function of gain, showing optimum confinement in the elongational axis at gain values ~5-6 V/$\mu$m. Solid lines are drawn to guide the eye. (b) Experimental frequency response ($\bullet$) of a trapped 190 nm polystyrene bead at different gain settings. Solid lines show Eq. \ref{eqn:freqdep} fitted to the data. $\omega_0$ values are highlighted with `$+$' markers. Data and fits have been vertically displaced for clarity. (c) Fitted values of $\gamma_{\text{eff}}$ ($\times$) show a linear dependence (linear fit shown by solid line) on gain setting, in agreement with $\gamma_{\text{eff}}\approx(c-k\tau)/m$.}
\end{figure}

To demonstrate the trapping capability of the electrokinetic stretcher, we trapped fluorescent polystyrene beads of varying sizes (190 nm, 800 nm, 930 nm) and double-stranded DNA molecules of varying lengths (48.5 kbp and 165.6 kbp). Fig. \ref{fig:trap} shows the trajectories of a 190 nm fluorescent bead and 165.6 kbp DNA molecule as they are 1) being actively trapped using a stretching voltage of \makebox{$(V_1+V_2)/2=$ 20 V}; and 2) escaping along the elongational axis. When the trap is active, the objects remain trapped for $>250$ s, with the standard deviation in displacements from the trap center $<1 \ \mu$m. On the other hand, untrapped objects escape along the elongational axis within $\sim1$ s.

Along the elongational ($x$) axis, we can estimate a spring constant (or trap stiffness)  $k=k_bT/\sigma^2$, where $k_b$ is the Boltzmann constant, $T$ is the temperature and $\sigma$ is the Gaussian width of the displacement distribution. For the examples presented in Fig. \ref{fig:trap}, we extract $k = 4$ nN/m. We highlight that tighter confinement can be achieved by optimizing the trap parameters, including the proportional gain setting and trap voltage (Fig. \ref{fig:trap}(a,b), inset).
Based on the minimum feedback latency of 10 ms, we can estimate the maximum diffusion coefficient of an object that the trap can confine to within a selected region of 500 nm, which is its minimum radius.\cite{FIELDS2010ABEL} During each feedback loop, the particle diffuses a root-mean-square (RMS) distance $d_{\text{rms},x}=(2D\tau)^{1/2}$, where $D=k_bT/6 \pi \eta r$ is the Stokes-Einstein diffusion coefficient, $\tau$ is the latency time, $\eta$ is the fluid viscosity and $r$ is the object radius. We thus estimate that an object can be confined to within 500 nm if \makebox{$r \approx 20$ nm}, at the system's limit of performance.

Among the adjustable trap parameters, the proportional gain affects the tightness of confinement in the elongational axis most significantly. In the compressional axis,  confinement improves with increasing stretching voltage instead, as shown in the inset of Fig. \ref{fig:trap}(a,b). The effect of adjusting gain can be seen from the positional RMS ($\sigma_x$ and $\sigma_y$) of a trapped 190 nm bead as gain is varied from 1-11 V/$\mu$m (Fig. \ref{fig:freqresp}(a)). While $\sigma_y$ does not show a clear trend with gain, there is an optimum gain value between 5 and 6 V/$\mu$m where $\sigma_x$ is smallest.

The frequency response of a $190$ nm bead is shown in Fig. \ref{fig:freqresp}(b). At the highest gain settings investigated, ringing occurs between 4 and 5 Hz, and is responsible for the upward trend in $\sigma_x$ with increasing gain. The observed behavior can be understood as akin to damped oscillation, due to the active feedback potential in the elongational axis, with motion driven by Brownian collisions. However, the feedback force on the bead is applied with a latency of a single frame time $\tau$. For such a time-delayed system, we thus write:
\begin{equation}
    m\ddot{x}(t) = -kx(t-\tau)-c\dot{x}(t)+\vec{F}(t),
    \label{eqn:timelag}
\end{equation}
where $t$ is time, $x$ is the displacement from the trap center, $k$ represents a spring constant dependent on the proportional gain selected for the feedback control, $c$ is a damping coefficient, $m$ is the object effective mass and $\vec{F}$ are the random impulsive forces due to Brownian collisions.
If the bead's velocity changes negligibly within a sufficiently small $\tau$, then we may approximate $x(t-\tau) \approx x(t)-\tau \dot{x}(t)$.\cite{19831943} Eq. \ref{eqn:timelag} acquires the familiar form of a damped harmonic oscillator:
\begin{align}
    m\ddot{x}(t) &= -kx(t)-(c-k\tau)\dot{x}(t)+\vec{F}(t),
\end{align}
for which there is a textbook frequency response:
\begin{align}
    \lvert x \rvert^2 &= \frac{A^2}{(\omega_0^2-\omega^2)^2 + (\gamma_{\text{eff}}\omega)^2}.
    \label{eqn:freqdep}
\end{align}
Here, $\gamma_{\text{eff}}=(c-k\tau)/m$ is an effective damping coefficient that depends on the spring constant (and consequently, gain value), $A^2=|\vec{F}|^2/m^2$ is a Brownian-driven amplitude that is frequency independent because the force spectrum due to Brownian impulses is white, and $\omega_0=\sqrt{k/m}$ is the natural frequency.

To validate our delayed-feedback oscillator model, we first fit Eq. \ref{eqn:freqdep} to the data from the highest gain setting (11.1 V/$\mu$m) to obtain values for $A$, $\omega_0$ and $\gamma_{\text{eff}}$. For the other gain settings, we fit Eq. \ref{eqn:freqdep} using the same value of $A$, and scale $\omega_0$ by the square root of the gain value, thereby constraining $\gamma_{\text{eff}}$ as the only degree of freedom during fitting. The fitted curves are shown in Fig. \ref{fig:freqresp}(b) and agree well with the data. In Fig. \ref{fig:freqresp}(c), we plot the fitted values of $\gamma_{\text{eff}}$ against the gain settings, and show a linear trend in agreement with the approximation made. The $x$-intercept in this plot suggests a sign flip of the $\gamma_{\text{eff}}$ term, corresponding to oscillation growth, rather than decay; that is, the object ultimately escapes the trap if the gain is set too high, as had been observed experimentally.

\subsection{High-throughput stretching of DNA molecules}

\begin{figure}[tb!]
\includegraphics[width=1\textwidth]{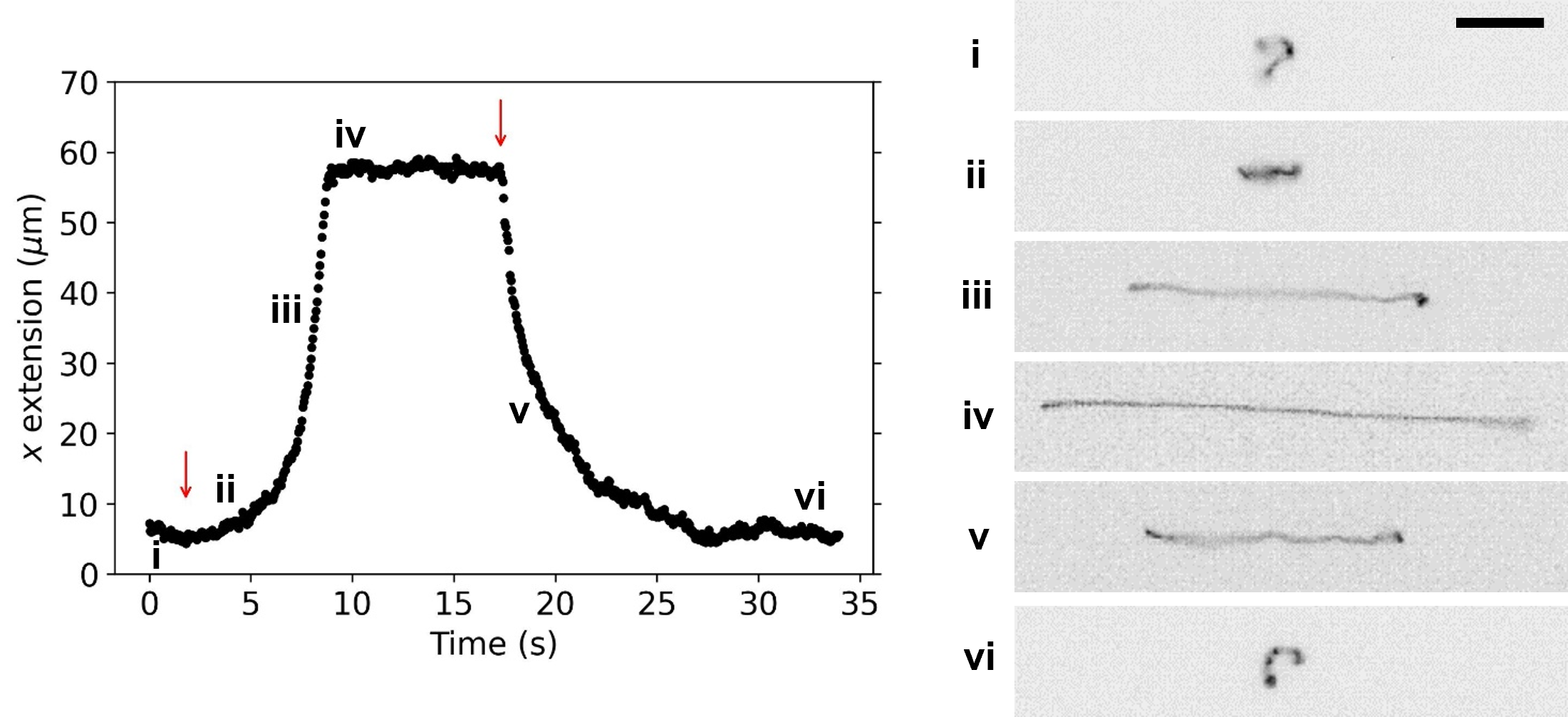}
\caption{Stretching and relaxing a double-stranded DNA molecule. Molecule length projected along elongational axis as a function of time (left) and series of images corresponding to the marked time points (right). The voltage was turned on and off as indicated by the red arrows. Scale bar represents 10 $\mu$m.}
\label{fig:stretch}
\end{figure}

\begin{figure*}[tb!]
\includegraphics[width = 0.75\textwidth]{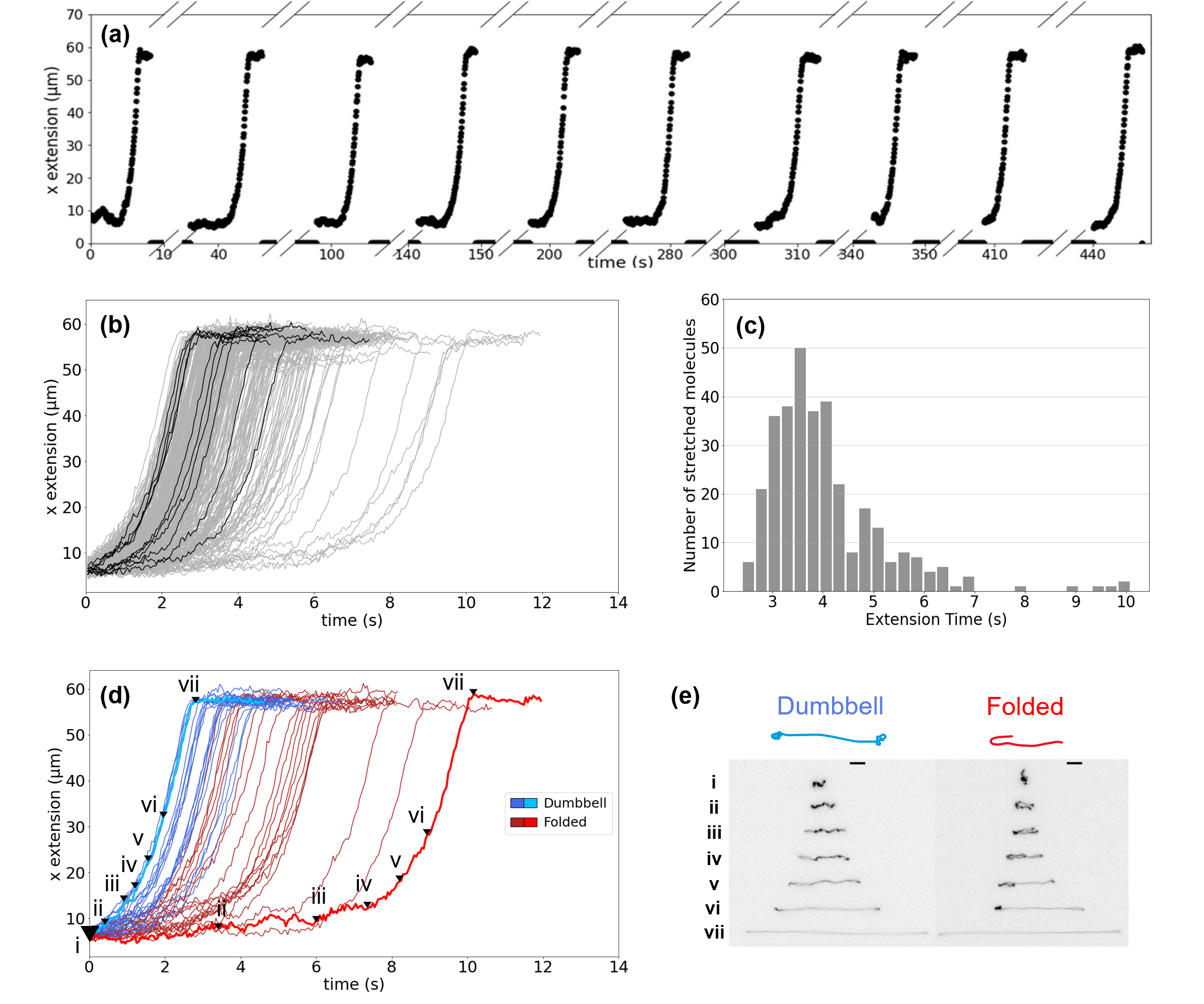}
  \caption{High-throughput measurements. (a) Ten DNA molecules trapped and stretched in series and fully autonomously by the platform; (b) Hundreds of stretching trajectories from data collected over 4 hours, at a rate of one single molecule stretch every $\sim$30 s. The ten molecules from (a) are highlighted in black; (c) Histogram of the extension time for the data shown in (b); (d) Selected stretching trajectories of molecules that adopted a transient dumbbell (blue) or folded (red) conformation; (e) Snapshots of a selected dumbbell (left) and folded (right) trajectory, at time points as labeled by markers in (d). Scale bars represent 5 $\mu$m.\label{fig:htsstretch}}
\end{figure*}

The elongational field generated in the cross-slot channel enables the stretching of individual molecules, which is of interest for biological and biotechnological applications.\cite{muller2017optical,mora2020nanomechanics} In this work, we demonstrate the stretching capability of our platform using DNA molecules. Fig. \ref{fig:stretch} shows the length of a DNA molecule projected along the elongational ($x$) axis as it undergoes stretching and subsequent relaxation at the stagnation point. The molecule is initially in a coiled configuration and begins to stretch when the field is turned on. Upon turning the field off, the molecule relaxes back into a coiled state. Note that the DNA can remain trapped in both axes even with an average trap voltage of 0 V as long as feedback correction is active, because whenever the object is displaced in the elongational axis (causing nonzero feedback), a passive trapping potential is also established in the compressional axis.

The platform was then programmed to perform the stretching of individual molecules in a self-driven, high-throughput manner (see Fig. S1 in the Supplementary Information for flowchart of algorithm). The algorithm involves three phases -- search, screen and stretch. In the search phase, the elongational field is turned on and molecules travel through the channel. Different molecules continuously enter the field of view of the camera, which is centered around the stagnation point. The program searches the image for a molecule to track and begins screening by checking the fluorescence intensity of the molecule. If this is greater than a pre-selected minimum intensity value, the molecule is determined to be intact and brought towards the stagnation point. Based on a running history of the molecule's position, the molecule is deemed to be trapped at the stagnation point when its RMS position falls below a threshold, upon which the stretch phase begins. The electric field is turned off for a period of time \makebox{(10 s)} -- this is because the molecule may stretch out en route to the stagnation point, so we allow time for the molecule to equilibrate at 0 V. The program then begins to record a video and the stretching voltage is turned back on to attain an elongational field. To determine if the molecule is in a fully stretched state, the program checks a short running history (1 s) of the molecule extension. If the range of values in this history is less than a certain threshold and the current extension is longer than a pre-selected minimum, the video recording is turned off and the program releases the molecule. The search--screen--stretch process then repeats with a new incoming molecule.

We demonstrate the automated, high-throughput stretching of DNA molecules using the electrokinetic stretcher (Fig. \ref{fig:htsstretch}). Within the first 8 minutes of starting the stretcher and without human intervention, the platform trapped and stretched 10 different DNA molecules in series, as seen in Fig. \ref{fig:htsstretch}(a). Over the next 4 hours, the platform recorded the trapping and stretching of 440 DNA molecules. Among these, roughly 75\% were chosen to be plotted in Figs. \ref{fig:htsstretch}(b--c) -- these molecules remained fully intact throughout the stretching process and were unperturbed by other molecules flowing past the stagnation point. 

Within the dataset, we observe diversity in the stretching trajectories that is consistent with that reported in the literature.\cite{perkins1997single,smith1998response} The heterogeneity in stretching dynamics arises from thermal fluctuations of the DNA molecules that lead to different random initial configurations when subjected to the field gradient. We can qualitatively classify the transient conformations of the molecules during extension. In Figs. \ref{fig:htsstretch}(d--e), we show sample trajectories that take on ``dumbbell'' and ``folded'' conformations. A ``dumbbell'' conformation occurs when the molecule unravels somewhat symmetrically, with the two coiled ends resembling a dumbbell; a ``folded'' conformation occurs when part of the molecule is folded over itself. A folded conformation is predisposed by thermal fluctuations at equilibrium that can cause both sides of the chain to lie on the same side of the molecule's center-of-mass when the electric field is turned on. Generally, we observe molecules that adopt a folded conformation tend to take a longer time to fully stretch, which is consistent with existing literature.\cite{perkins1997single,smith1998response} However, this is a coarse classification process; we observe from Fig. \ref{fig:htsstretch}(d) that a molecule in a dumbbell conformation does not necessarily stretch more rapidly than a molecule in a folded conformation. We envision more quantitative and fine-tuned analysis can be performed on such a dataset using data-driven learning approaches to gain further insight into the dynamics of the polymer stretching process. Importantly, this can be facilitated by the platform developed in this work, which allows for the collection of a large amount of video data with minimal human intervention.

\section{Discussion}

\begin{figure}[tb]
\includegraphics[width = 1\textwidth]{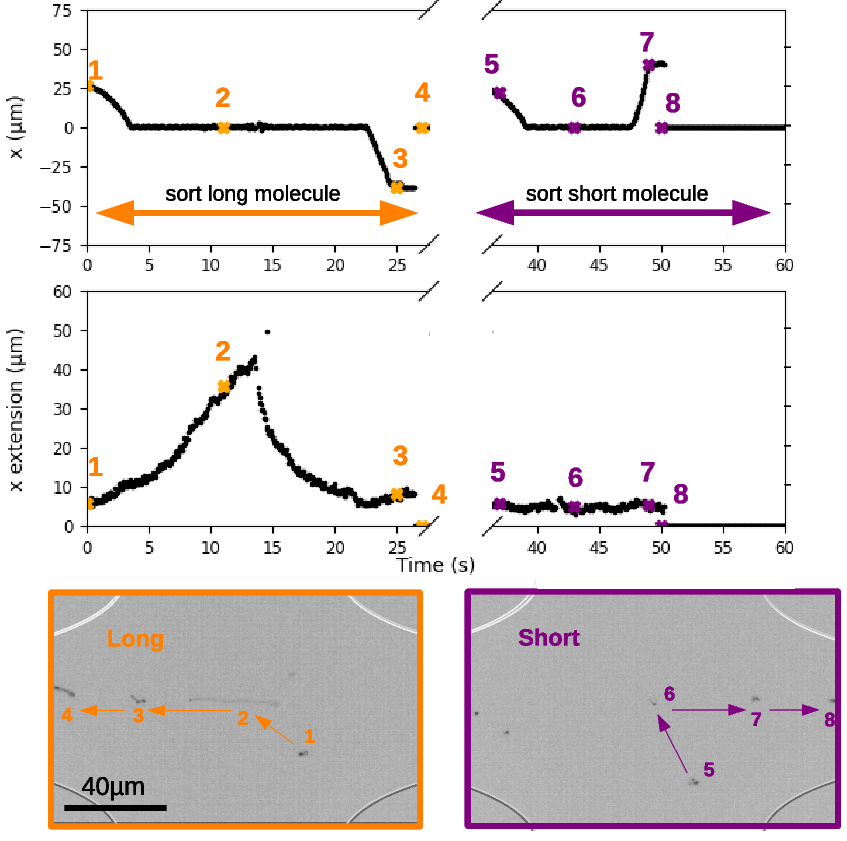}
 \caption{Sequential trapping and sorting of two DNA molecules. Top: Location along the $x$ axis of a long (orange) and a short (purple) DNA molecule. Middle: Variation in extension length of the DNA molecules throughout the sorting process. Bottom: Overlay images showing the displacement and morphology of the two molecules in the center of the microfluidic channel. For clarity, the images' intensity gradients have been inverted.}\label{fig:sort}
\end{figure}

The capabilities of this platform are well-suited for the high-throughput observation and stretching of single molecules and particles. As trapping and stretching can be automated with minimal human intervention, large datasets containing a range of information about individual molecules or particles can be collected at the single-molecule level, and subsequently studied using big data analysis techniques. 

Apart from trapping and stretching objects for long observation times, the electrokinetic stretcher presented in this work can also enable other types of contact-free manipulation of micro- and nanoscale objects. For example, the stretcher can be used to perform size-based sorting of molecules. This is demonstrated using a mixture of 48.5 kbp and 165.6 kbp DNA molecules, as illustrated in Fig. \ref{fig:sort}. In the example presented, both long and short molecules are captured around $x=25\ \mu$m from the center of the trap, before being brought to the stagnation point. The molecules remain trapped at $x\sim 0\ \mu$m while they are stretched and classified as long or short based on their maximum projected length along the elongational axis (40$\ \mu$m and 10$\ \mu$m for the long and short DNA, respectively). After classification, the trapping voltage is turned off and the molecules return to a coiled configuration. The molecules are then moved towards either the left or right reservoir ($x\sim -30\ \mu$m and $x\sim 40\ \mu$m, respectively). Finally, the system releases the classified molecule and awaits a new molecule to capture and sort. The capability to sort nanoscale objects is highly sought after in the field of nanomaterials, and we can envision our platform as a high-throughput sorter for nanoscale or microscale objects. We highlight that the flexibility of the platform opens up a multitude of possibilities for downstream applications. For example, while we have demonstrated sorting of DNA molecules based on size, it is equally feasible to sort objects based on other observable features, such as shape or fluorescence intensity.

Using electrokinesis (\textit{i.e.} electrophoresis and/or electroosmosis) as the trapping mechanism presents several advantages over other microfluidic-compatible object manipulation techniques. Compared to the radiation pressure in optical tweezing, electrokinetic forces scale much more favorably with decreasing size.\cite{FIELDS2010ABEL} This enables the trapping of smaller objects (< 100 nm),\cite{kayci2014nanodiamond,Dissanayaka20193Dfeedback,ropp2013QDnanorod} and even single fluorophores.\cite{fields2011electrokinetic} Hydrodynamic microflows have also been used in the Stokes trap,\cite{tanyeri2010hydrodynamic,shenoy2016stokes} but the hydrostatic pressure-driven flow profiles necessitate the use of deep channels to avoid the sharp velocity gradient resulting from the no-slip condition at channel walls. In contrast, electrokinetic flow profiles are flat,\cite{Probst2012FlowControl} allowing the use of channels with heights $\sim 1$ $\mu$m, confining the particle vertically to within the focal depth of high numerical aperture microscope objectives. Another more practical advantage is the facile implementation as we have presented here -- besides the initial relief mold fabrication, no other micro-fabrication steps were necessary. We find the fields to be robust to variations in reservoir and electrode positions, showing highly symmetric flow patterns even with hand-aligned reservoirs and electrodes.

While various forms of microfluidic electrokinetic traps have been reported in the literature, such as the ABEL trap \cite{cohen2005method} and the aqueous Paul trap, \cite{guan2011paul} we advance the technology by actively stretching the trapped molecule, driving it far from equilibrium. This is made possible by: first, a controllable elongational field; and second, the tracking algorithm that is capable of tracking extended, deformable objects with sufficiently low latency. While density-based clustering algorithms are familiar in data analysis and machine learning, we report here the use of clustering as a means of real-time localization, applicable to deformable as well as point-like objects.

\section{Conclusions}
In this work, we have presented an automated, high-throughput method of trapping and stretching individual micro- and nanoscale objects in solution using an electrokinetic stretcher. The platform operates on a feedback control system that calculates and maintains the center-of-mass of the object at the stagnation point of a planar elongational electric field, even if the object's shape is constantly evolving. With our platform, we have managed to track and confine single submicron polystyrene beads and double-stranded DNA molecules for periods exceeding 250 s, with a standard deviation in displacement from the trap center of less than 1 ${\mu}$m without viscosity engineering. We have also demonstrated the high-throughput stretching capabilities of our platform by stretching $>$400 DNA molecules over 4 hours without human intervention. 

The trap-and-stretch capability is relevant scientifically (e.g. biopolymer physics \cite{smith1998response,randall2005dna,gross2011quantifying,klotz2018motion}) and technologically (e.g. sequencing and mapping \cite{lam2012genome,keyser2011controlling}). Our demonstration of self-driven, high-throughput stretching of DNA molecules shows the robustness, speed and reproducibility of the platform, which opens up fresh opportunities at the intersection of biomolecule physics and data sciences. Apart from stretching, other types of contact-free manipulation of nanoscale objects are also possible, an example of which is size-based sorting of molecules. Overall, the platform provides an automated method for the tracking and versatile manipulation of micro- and nanoscale objects, empowering static and dynamic studies of single molecules and biological systems in real time.

\section*{Acknowledgments}
\noindent
The authors acknowledge funding from the Accelerated Materials Development for Manufacturing Program via the AME Programmatic Fund by the Agency for Science, Technology and Research (A*STAR) under Grant No. A1898b0043 and Career Development Fund by A*STAR under Grant No. C222812024. K.H. also acknowledges funding from the NRF Fellowship NRF-NRFF13-2021-0011.

\section*{Author Declarations}
\subsection*{Conflict of Interest}
\noindent
The authors have no conflicts to disclose.

\subsection*{Author Contributions}
\noindent
\textbf{Beatrice W. Soh:} Conceptualization; investigation; formal analysis; visualization; writing - original draft; writing - review and editing. \textbf{Zi-En Ooi:} Conceptualization; software; investigation; formal analysis; visualization; writing - original draft; writing - review and editing. \textbf{Eleonore Vissol-Gaudin:} Investigation; formal analysis; visualization; writing - original draft; writing - review and editing. \textbf{Chang Jie Leong:} Investigation; visualization; writing - original draft; writing - review and editing. \textbf{Kedar Hippalgaonkar:} Funding acquisition; writing - review and editing.

\bibliographystyle{unsrt}
\bibliography{Electrophoretic_DNA_Trap}

\end{document}


\doublespacing	
\title{\textbf{Supplementary Information for Automated Electrokinetic Stretcher for Manipulating Nanomaterials}}
\author[1]{Beatrice W. Soh$^*$}
\author[1]{Zi-En Ooi}
\author[2]{Eleonore Vissol-Gaudin}
\author[1]{Chang Jie Leong}
\author[1,2]{Kedar Hippalgaonkar\thanks{beatrice\_soh@imre.a-star.edu.sg and kedar@ntu.edu.sg}} \vspace{1cm}
\affil[1]{Institute of Materials Research and Engineering, Agency for Science, Technology and Research, Singapore}
\affil[2]{Department of Materials Science and Engineering, Nanyang Technological University, Singapore}
\date{}

\maketitle
\clearpage
\section*{Flowchart for high-throughput stretching algorithm}
\begin{figure*}[h!]
\includegraphics[width=\textwidth]{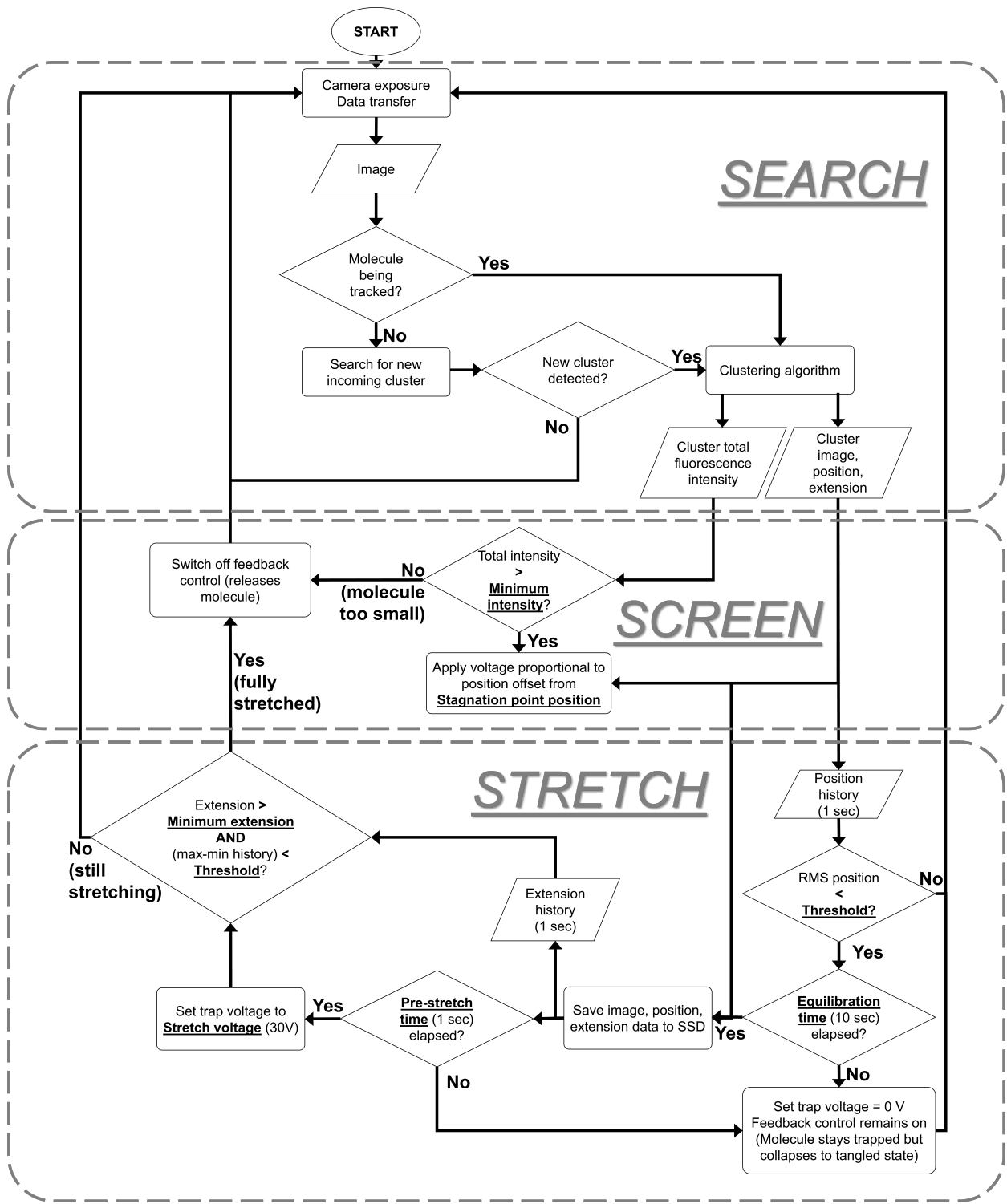}
\caption{Flowchart summary of the automated DNA stretching program.}
\label{fig:flowchart}
\end{figure*}